\documentclass[
reprint,
superscriptaddress,
amsmath,
amssymb,
aps,
prl,
]{revtex4-1}

\usepackage[pdftex]{graphicx}
\usepackage{comment}
\usepackage{url}
\usepackage[labelformat=empty,labelsep=none]{caption}
\usepackage{bm}
\usepackage{comment}

\begin{document}

\title{Mapping Observation Project of High-Energy Phenomena\\during Winter Thunderstorms in Japan}

\author{Yuuki~Wada}
\affiliation{Department of Physics, Graduate School of Science, The University of Tokyo, 7-3-1 Hongo, Bunkyo-ku, Tokyo 113-0033, Japan}
\affiliation{High Energy Astrophysics Laboratory, Nishina Center for Accelerator-Based Science, RIKEN, 2-1 Hirosawa, Wako, Saitama 351-0198, Japan}
\author{Teruaki~Enoto}
\affiliation{The Hakubi Center for Advanced Research and Department of Astronomy, Kyoto University, Kitashirakawa Oiwake-cho, Sakyo-ku, Kyoto 606-8502, Japan}
\author{Yoshihiro~Furuta}
\affiliation{Collaborative Laboratories for Advanced Decommissioning Science, Japan Atomic Energy Agency, 2-4 Shirane Shirakata, Tokai-mura, Naka-gun, Ibaraki 319-1195, Japan}
\author{Kazuhiro~Nakazawa}
\affiliation{Kobayashi-Maskawa Institute for the Origin of Particles and the Universe, Nagoya University, Furo-cho, Chikusa-ku, Nagoya, Aichi 464-8601, Japan}
\author{Takayuki~Yuasa}
\affiliation{Block 4B, Boon Tiong Road, Singapore 165004, Singapore}
\author{Takahiro~Matsumoto}
\affiliation{Department of Physics, Graduate School of Science, The University of Tokyo, 7-3-1 Hongo, Bunkyo-ku, Tokyo 113-0033, Japan}
\author{Daigo~Umemoto}
\affiliation{Discrete Event Simulation Research Team, Center for Computational Science, RIKEN, 7-1-26 Minatojima-minami-machi, Chuo-ku, Kobe, Hyogo, 650-0047 Japan}
\author{Kazuo~makishima}
\affiliation{High Energy Astrophysics Laboratory, Nishina Center for Accelerator-Based Science, RIKEN, 2-1 Hirosawa, Wako, Saitama 351-0198, Japan}
\affiliation{Department of Physics, Graduate School of Science, The University of Tokyo, 7-3-1 Hongo, Bunkyo-ku, Tokyo 113-0033, Japan}
\affiliation{Kavli Institute for the Physics and Mathematics of the Universe, The University of Tokyo, 5-1-5 Kashiwa-no-ha, Kashiwa, Chiba 277-8683, Japan}
\author{Harufumi~Tsuchiya}
\affiliation{Nuclear Science and Engineering Center, Japan Atomic Energy Agency, 2-4 Shirane Shirakata, Tokai-mura, Naka-gun, Ibaraki 319-1195, Japan}
\author{the GROWTH collaboration}

\begin{abstract}
The Gamma-Ray Observation of Winter Thunderclouds (GROWTH) collaboration 
	has been performing observation campaigns of high-energy radiation in coastal areas of Japan Sea.
	Winter thunderstorms in Japan have unique characteristics such as frequent positive-polarity discharges,
	large discharge current, and low cloud bases. These features allow us to observe 
	both long-duration gamma-ray bursts and lightning-triggered short-duration bursts at sea level. 
	In 2015, we started a mapping observation project using multiple detectors at several new observation sites. 
	We have developed brand-new portable gamma-ray detectors and deployed in the Kanazawa and Komatsu areas as well as the existing site at Kashiwazaki.
	During three winter seasons from 2015, we have detected 27 long-duration bursts and 8 short-duration bursts. 
	The improved observation network in Kashiwazaki enables us to discover that the short-duration bursts 
	are attributed to atmospheric photonuclear reactions triggered by a downward terrestrial gamma-ray flash.
	Collaborating with electric-field and radio-band measurements, we have also revealed a relation between abrupt termination of a long-duration burst and a lightning discharge. 
	We demonstrate that the mapping observation project has been providing us clues to understand high-energy atmospheric phenomena associated with thunderstorm activities.
\end{abstract}

\maketitle


\section{introduction}
Recent discoveries of high-energy phenomena associated with thunderstorm activities have been proving that
	thunderclouds and lightning discharges can be powerful electron accelerators.
	From space, terrestrial gamma-ray flashes (TGFs) were first discovered by Compton Gamma-Ray Observatory\cite{Fishman_1994}, 
	then have been observed by gamma-ray astronomy satellites such as RHESSI\cite{Smith_2005}, AGILE\cite{Marisaldi_2015} and Fermi\cite{Briggs_2011}.
	They last for a few hundred microseconds to several milliseconds, and their photon energy extends beyond 20~MeV.
	In addition, similar lightning-associated events at ground-level have been also detected by mountain-top experiments\cite{Moore_2001,Colalillo_2017,Abbasi_2017,Abbasi_2018}
	and by rocket-triggered lightning experiments\cite{Dwyer_2003a,Hare_2016}. They are referred to as ``downward terrestrial gamma-ray flashes''.
	In contrast, long-lasting radiation enhancements from thunderclouds have been also detected
	by airborne\cite{McCarthy_1985,Eack_1996,Kelley_2015,Kochkin_2017}, 
	mountain-top\cite{Brunetti_2000,Torii_2009,Tsuchiya_2009,Tsuchiya_2012,Chilingarian_2010,Chilingarian_2011,Chilingarian_2016}, 
	and sea-level measurements\cite{Torii_2002,Torii_2011,Tsuchiya_2007,Tsuchiya_2011,Kuroda_2016}. 
	They are called long bursts\cite{Torii_2011}, gamma-ray glows\cite{Kelley_2015}, and thunderstorm ground enhancements especially when detected by on-ground experiments\cite{Chilingarian_2011}.
	They have second- to minute-order duration which is much longer than TGFs, and their photon energy can also reach a few tens of MeV\cite{Tsuchiya_2011}.
	The long-lasting emissions often precede and sometimes terminate 
	with lightning discharges\cite{McCarthy_1985,Eack_1996,Alexeenko_2002,Tsuchiya_2007,Tsuchiya_2013,Kelley_2015,Chilingarian_2015,Chilingarian_2017}.
	
These atmospheric high-energy phenomena are thought to be bremsstrahlung of electrons accelerated in strong electric fields of lightning and thunderclouds.
	Based on Wilson's runaway electron hypothesis\cite{Wilson_1925}, Grevich et al.\cite{Gurevich_1992} proposed relativistic runaway electron avalanches (RREA).
	When thunderstorms have strong electric fields (e.g. more than 284~kV/m at standard temperature and pressure, derived by a simulation of Dwyer\cite{Dwyer_2004}.), 
	energetic seed electrons, whose energy is more than a few hundreds of keV, are accelerated and exponentially multiplied.
	RREA is thought to be the most plausible model for these high-energy phenomena associated with thunderstorm activities.
	In addition, the relativistic feedback model was introduced by Dwyer\cite{Dwyer_2012} to explain the brightness of TGFs.
	
Winter thunderstorms in Japan are ideal targets for observations of atmospheric high-energy phenomena.
	Long bursts in Japanese winter thunderstorms were first discovered by radiation monitoring stations in nuclear power plants\cite{Torii_2002},
	and have been observed by sea-level measurements\cite{Torii_2002,Torii_2011,Tsuchiya_2007,Tsuchiya_2011,Kuroda_2016}.
	Winter thunderstorms have unique features comparing to summer ones such as high-current discharges, 
	a large proportion of positive-polarity discharges and upward leaders, and lower cloud bases\cite{Goto_Narita_1992,Rakov_2003}.
	In usual, long bursts in summer thunderstorms can be hardly detected at sea-level 
	because their charged region is located typically above 3~km altitude, 
	which is too high for gamma-rays of MeVs to penetrate toward the ground.
	On the other hand, lower cloud bases of winter thunderstorms, typically less then 1~km, allow gamma rays to reach the sea level.
	
In order to investigate high-energy phenomena in winter thunderstorms,
	we started the GROWTH (Gamma-Ray Observation of Winter Thunderclouds) experiment in 2006.
	Radiation monitors were deployed at Kashiwazaki-Kariwa Nuclear Power Plant in Niigata Prefecture, Japan.
	Coastal areas of japan Sea, including the Kashiwazaki site, often encounter active thunderstorms during every winter season.
	We have observed two types of energetic phenomena. Long bursts, as referred above, 
	are minute-order bremsstrahlung emissions from thunderclouds, apparently not associated with lightning\cite{Tsuchiya_2007,Tsuchiya_2011,Tsuchiya_2013}.
	Tsuchiya et al.\cite{Tsuchiya_2007} revealed that long bursts originate from bremsstrahlung of electrons accelerated beyond 10~MeV in winter thunderclouds. 
	In contrast, we have also detected short-duration radiation bursts called ``short bursts''\cite{Umemoto_2016} coinciding with lightning discharges.
	Short bursts has duration of a few hundred milliseconds, which is shorter than long bursts, but longer than TGFs.
	Typical count-rate histories and energy spectra of the two phenomena are presented in Figure~1.

\begin{figure*}[hbt]
	\begin{center}
	\includegraphics[width=0.8\hsize]{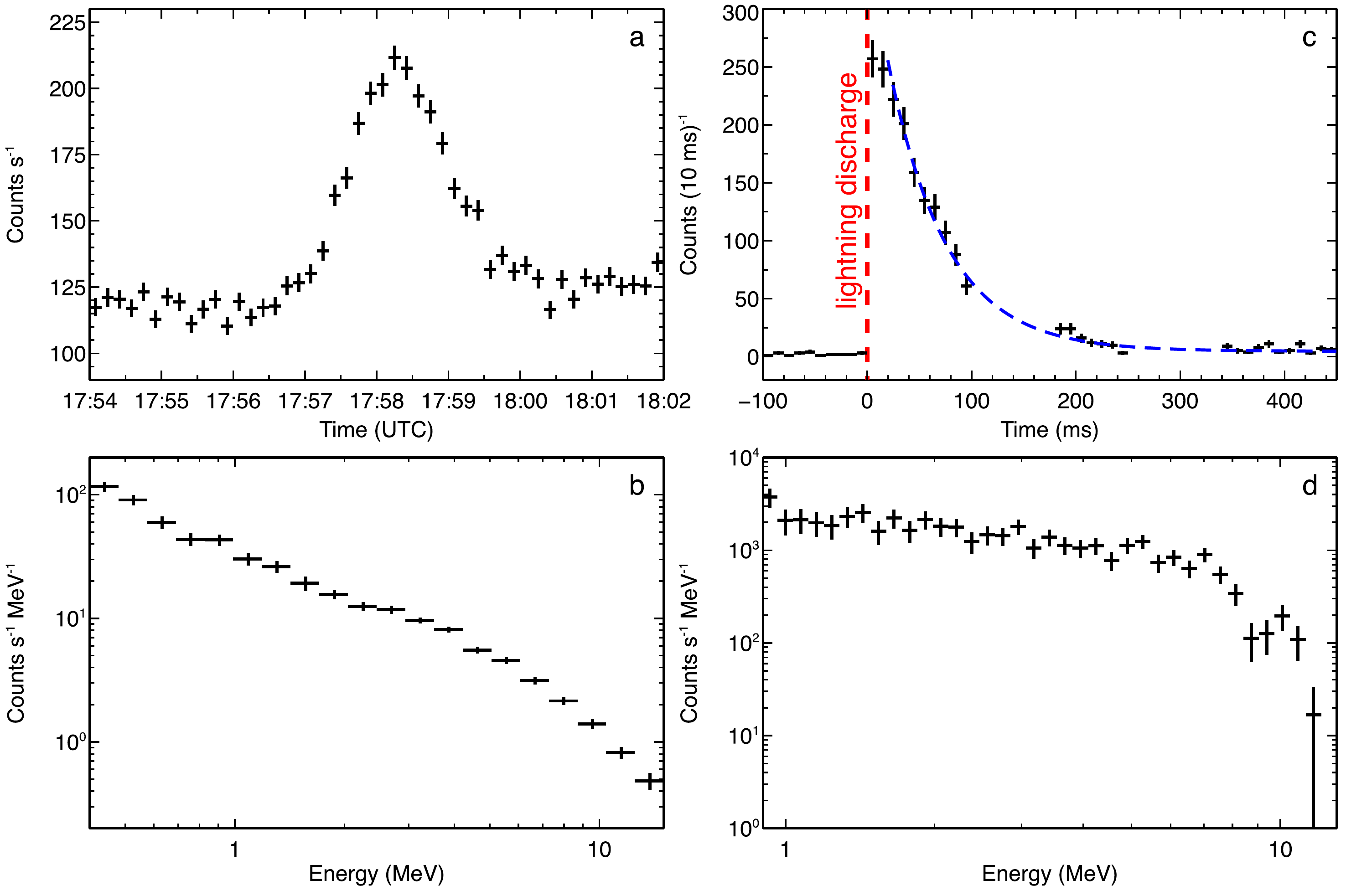}
	\caption{
	Figure 1: Typical time series of count rates and spectra of long bursts (a and b) and short bursts (c and d). 
	(a) A 10-sec-binned count-rate history of a long burst in 0.7-15.0 MeV, observed in Komatsu on 8th December 2016.
	(b) An energy spectrum of the long burst shown in panal a, extracted from 17:56:30--17:57:30 (UTC). 
	Detector responses remain unremoved.
	(c) A count-rate history of a short burst in 0.35--20.0 MeV, observed in Kashiwazaki on 6th February 2017. 
	The origin of the X axis and the red-dashed line show the timing of a lightning discharge at 08:34:06.
	The best-fit exponential function to the count-rate history are overlaid with the blue-dashed line.
	(d) An energy spectrum of the short burst in panel b, extracted from 40--100~ms.
	}
	\end{center}
\end{figure*}

As observational results of long bursts are accumulated, several important questions to be answered are raised:
	\begin{itemize}
	\setlength{\itemsep}{0pt}
	\setlength{\parskip}{0pt}
	\setlength{\labelsep}{5pt}
	\setlength{\leftmargin}{10pt}
	\item How long bursts emerge, develop, and terminate?
	\item How thunderclouds keep highly electrified region responsible for electron acceleration?
	\item How large energy thunderclouds release by emitting high-energy photons?
	\end{itemize}
	In addition, what causes short bursts was completely missing, which is addressed later (see setion "Inter-pretation of short bursts").
	To answer these questions, we started a mapping observation project with multiple observation sites.
	The project was launched in 2015 with 2 portable detectors, 
	and is expanding the number of detectors in coastal areas of Japan Sea.
	The project aims to detect long bursts and short bursts with multiple detectors,
	and to measure spatial flux distribution, spectra, and temporal flux variations. 
	In the present paper, we introduce our detectors dedicated to the project,
	and high-lights of observational results in 2015-2017 winter seasons.

\section{Instrumentation}
\subsection{Development of portable detector}

The mapping observation project requires portable detector system dedicated to outdoor observation of gamma-rays.
	We have been deploying detectors mainly in urban areas because it is easy to obtain power sources.
	In this case, detectors should be compact for limited installation spaces on rooftops.
	In addition, to deploy more than 10 detectors,  the detector system should be portable, easy to handle, 
	easy to assemble, and suitable for mass production.
	Therefore, we employed a simple configuration consisting of a main scintillation crystal coupled with photomultipliers (PMTs), 
	data acquisition (DAQ) system, and telecommunication system.
	
Scintillation crystals are utilized as the main detection component. To detect gamma rays of more than 10~MeV, 
	we employ inorganic scintillators such as Be$_{4}$Ge$_{3}$O$_{12}$ (BGO), sodium iodide (NaI), and cesium iodide (CsI) crystals.
	For example, BGO crystals with 2.5 cm thickness can interact with $\sim$50\% of 10~MeV gamma rays.
	Light yields form these scintillation crystals are read by PMTs.
	Although we utilize various crystals and PMTs for mass production, 
	our popular configurations are 25$\times$8$\times$2.5~cm$^{3}$ BGO crystals coupled with Hamamatsu R1924 PMTs, 
	and 30$\times$5$\times$5~cm$^{3}$ CsI crystals with Hamamatsu R6231 PMTs.

Telecommunications are performed via mobile phone network.
	A mobile router is employed to connect the DAQ system to the cellular network such as the Long Term Evolution network in Japan.
	The DAQ system continually send telemetries of operation status, temperature of the system, electricity consumption, and so on.
	Due to a limited amount of data transfer, all data cannot be sent in real time.
	Instead, we can download required data (e.g. during thunderstorms) on demand.
 
 These components are packed in a waterproof box BCAR453520T (Takachi Electronics Enclosure), 
 	whose size is 45~cm (width) $\times$ 35~cm (length) $\times$ 20~cm (height).
	The inside and outside of the detector are presented in Figure~2. Electricity is supplied via a waterproof cable.
	A typical weight of the whole system is $\sim$15~kg, depending on scintillation crystals.
	The system is fixed to concrete blocks or building facilities in order to prevent it from flying away due to severe winter thunderstorms.

\begin{figure*}[ht]
	\begin{center}
	\includegraphics[width=0.8\hsize]{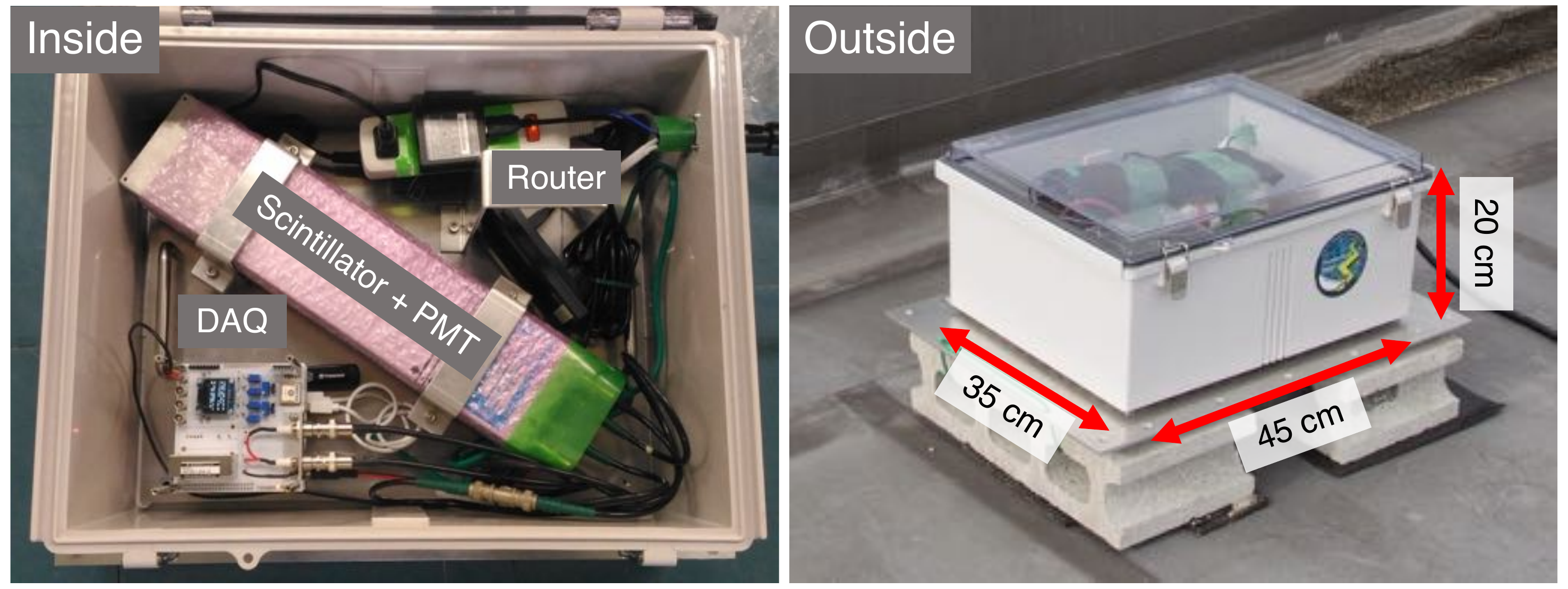}
	\caption{
	Figure2: The inside (left) and outside (right) of the detector.
	}
	\end{center}
\end{figure*}

\subsection{Data acquisition system}
It is necessary to develop a compact DAQ system for portable detectors.
	In our configuration, scintillation crystals and DAQ system can be made small.
	However, the size of main scintillation crystals, which affects the sensitivity to gamma rays, cannot be smaller than necessary.
	Therefore, we focused on developing a compact new DAQ system dedicated to the mapping observation project. 
	The DAQ system consists of three components: GROWTH FPGA/ADC board, GROWTH daughter board, and Raspberry Pi.
	The photographs and block diagram of the DAQ system is presented in Figure~3.

\begin{figure*}[ht]
	\begin{center}
	\includegraphics[width=0.8\hsize]{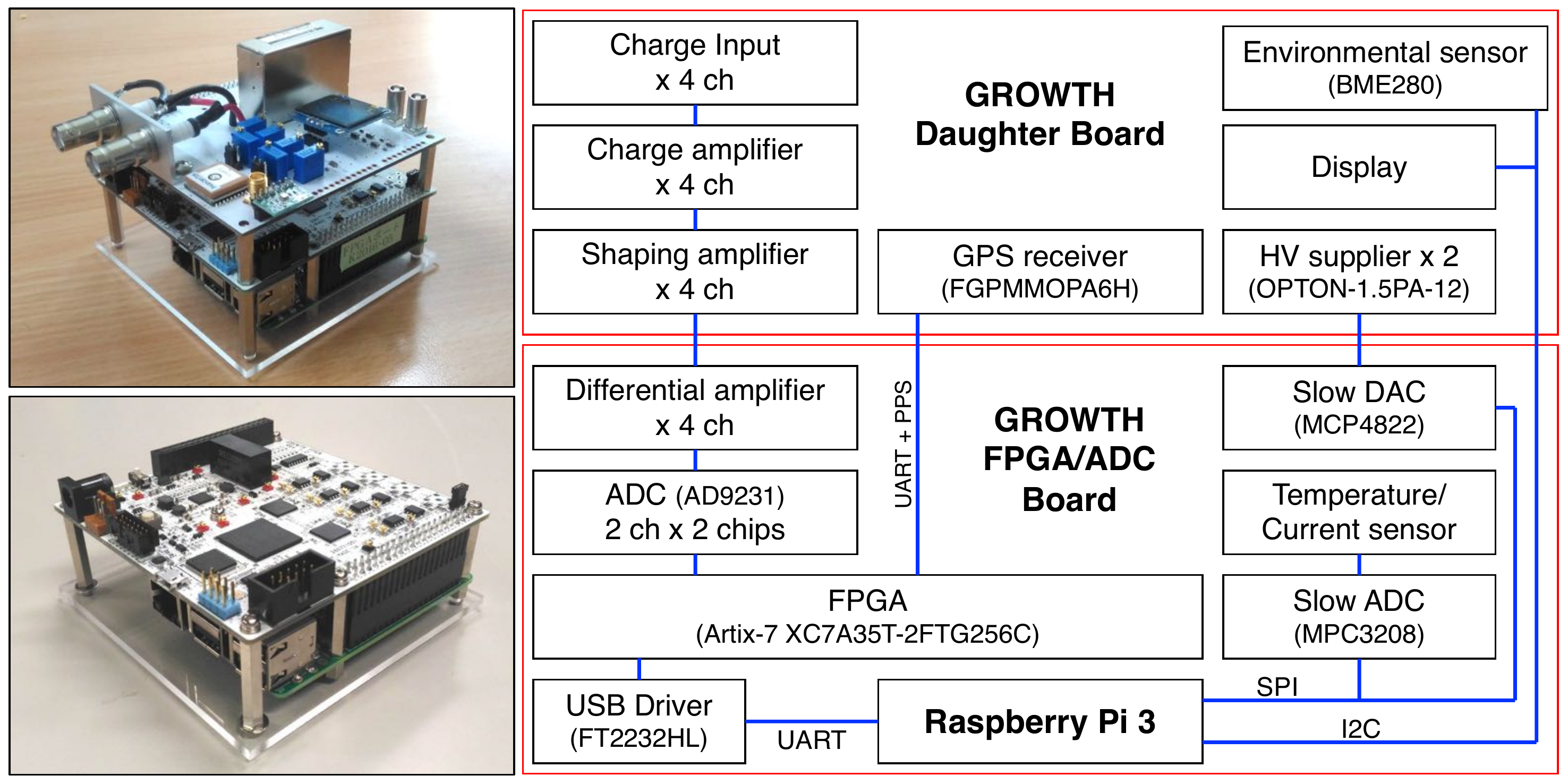}
	\caption{
	Figure 3: Photographs of the DAQ system with (left top) and without (left bottom) the GROWTH daughter board, and a block diagram (right).
	}
	\end{center}
\end{figure*}
	
The GROWTH FPGA/ADC board has been developed in cooperation with Shimafuji Electronics Co., Ltd.
	It is a general-purpose analog-to-digital converting board with a field programming gate array (FPGA).
	It has 4-channel analog inputs with $-$5~V to 5~V coverage
	The input signals are at first buffered by differential amplifiers, then sampled by 12-bit ADC chips (Analog Devices AD9231). 
	The ADC chip can be operated with up to 65~MHz. We employ the 50~MHz sampling rate for the experiment.
	The converted signals are processed by FPGA (XILINX Artix-7 XC7A35T-2FTG256C).
	We employed a self-trigger system. Once the input signal gets over the threshold, 
	FPGA extracts maximum and minimum values and arrival time of the sampled data during a gate time.
	The extracted information from sampled waveform is sent via a USB-driving chip (FTDI FT2232HL) 
	to Raspberry Pi 3 by the Universal Asynchronous Receiver/Transmitter (UART) interface.
	The gate time and trigger thresholds are also modifiable via Raspberry Pi.
	In addition, a slow ADC chip (Microchip Technology MPC3208) connected to temperature and current sensors, 
	and a slow digital-to-analog convertor (DAC) chip (Microchip Technology  MPC4822) are onboard.
	These slow ADC and DAC are controlled by Raspberry Pi via Serial Peripheral Interface.
	The GROWTH FPGA/ADC board is also connected with Raspberry Pi 
	via 2$\times$20-pin GPIO (general purpose input/output), besides the UART interface.
	The FPGA/ADC board is powered by DC12~V input. 
	
Since the GROWTH FPGA/ADC board is a general-purpose ADC board, it has no amplifier circuits 
	dedicated to e.g. PMT and silicon photomultiplier readouts.
	Instead, the board can be connected to a daughter board designed for a certain purpose via a 2$\times$20-pin connector.
	As described in the next paragraph, daughter boards are expected to have charge amplifiers, waveform-shaping amplifiers, 
	high-voltage suppliers, temperature sensors, and so on.
	Also, a global positioning system (GPS) receiver on daughter boards can be connected to the ADC/FPGA board.
	If the receiver obtains GPS signals properly, accurate absolute timing (better than 1~$\mu$s) is assigned to the digitized waveform by FPGA. 
	Otherwise the absolute timing is assigned by internal clock of Raspberry Pi with an accuracy of $\sim$1~s.

We also developed the GROWTH daughter board to read PMT outputs.
	The daughter board has 4-channel charge amplifiers and shaping amplifiers.
	These two amplifiers have time constants of 10~$\mu$s and 2~$\mu$s, respectively.
	The amplified signals are sent to the FPGA/ADC board via the 2$\times$20-pin connector, then sampled by the ADC chips.
	A GPS receiver (Global Top FGPMMOPA6H) is onboard and connected to FPGA. A GPS antenna can be connected to the daughter board via SMA terminal.
	The daughter board has also a module-type high-voltage suppliers for PMTs (Matsusada Precision OPTON-1.5PA/NA-12) which can supply 0--1500~V.
	The high-voltage supplier accepts reference voltage to control output voltage.
	In our case, we utilize the slow DAC on the FPGA/ADC board to generate the reference voltage, 
	thus the output voltage can be controlled by Raspberry Pi. PMTs can be connected to the high-voltage suppliers via SHV connectors.
	In addition, the daughter board has a small display (OLED SSD1306) and an environmental sensor to measure temperature, humidity, and atmospheric pressure.
	They are connected to Raspberry Pi through the FPGA/ADC board and controlled by the Inter-Integrated Circuit interface.
	The size of the assembled DAQ system is 9.5~cm (width) $\times$ 9.5~cm (length) $\times$ 10.3~cm (height) including the daughter board and HV suppliers.
	Total electricity consumption including Raspberry Pi is 7~W.

\subsection{Deployment}
Since the launch of the mapping observation project in 2015, we are expanding the number of detectors.
	In 2017--2018 winter season, we had five observation sites 
	in Kanazawa, Komatsu, Toyama, Suzu, and Kashiwazaki with 16 detectors in total.
	Figure~4 presents observation sites in the 2017--2018 season.
	All the observation sites are located in coastal areas of Japan Sea.
	Suzu and Toyama sites are at Universities. Kanazawa and Komatsu sites consist of a university, local high schools, and a science museum.
	The Kashiwazaki site is in Kashiwazaki-Kariwa Nuclear Power Plant, where we have been performing the GROWTH experiments since 2006, 
	and updated with 4 detectors in 2016.
	
\begin{figure}[ht]
	\begin{center}
	\includegraphics[width=\hsize]{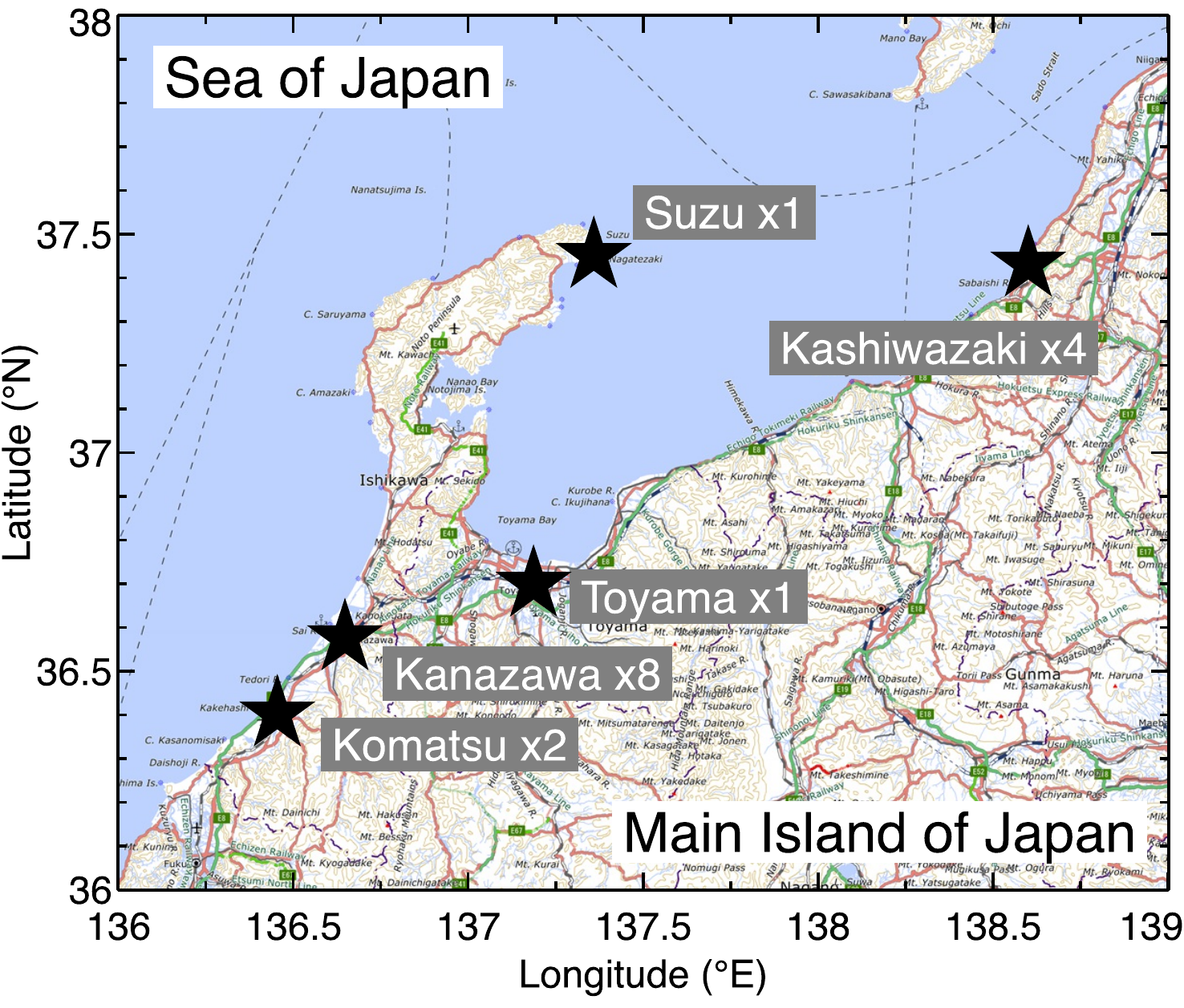}
	\caption{
	Figure 4: The observation sites in the 2016-2017 winter season.
	}
	\end{center}
\end{figure}

\subsection{Calibration}
We perform timing and energy calibration for obtained data.
	Timing of each photon event is assigned by GPS signals. Successfully-received GPS signals are confirmed to give absolute timing better than 1~$\mu$s.
	The energy calibration is performed by using persistent background radiation such as the 1.46~MeV line of $^{40}$K and the 2.61~MeV line of $^{208}$Tl.
	In addition, lines from $^{214}$Bi are also utilized to estimate the accuracy of the energy calibration.
	Figure~5 presents background spectra obtained by a detector at Kanazawa University in March 2018.
	During raining, count rates below 3~MeV because $^{214}$Bi, which is a daughter product of $^{222}$Rn, resides in rain drops, then falls onto the ground	.
	In this case, the 0.609~MeV line of $^{214}$Bi is suitable to investigate the calibration accuracy.
	When the energy is calibrated by a linear function derived from 1.46~MeV and 2.61~MeV background lines, 
	the accuracy of the energy calibration is less than 2\% at 0.609~MeV.
	Since BGO scintillation crystals have a temperature dependence on light yields, these calibration procedures are performed for every 30 minutes.
	Whereas, the procedures for NaI and CsI crystals are performed dairy due to the low temperature dependences.
	
\begin{figure}[ht]
	\begin{center}
	\includegraphics[width=\hsize]{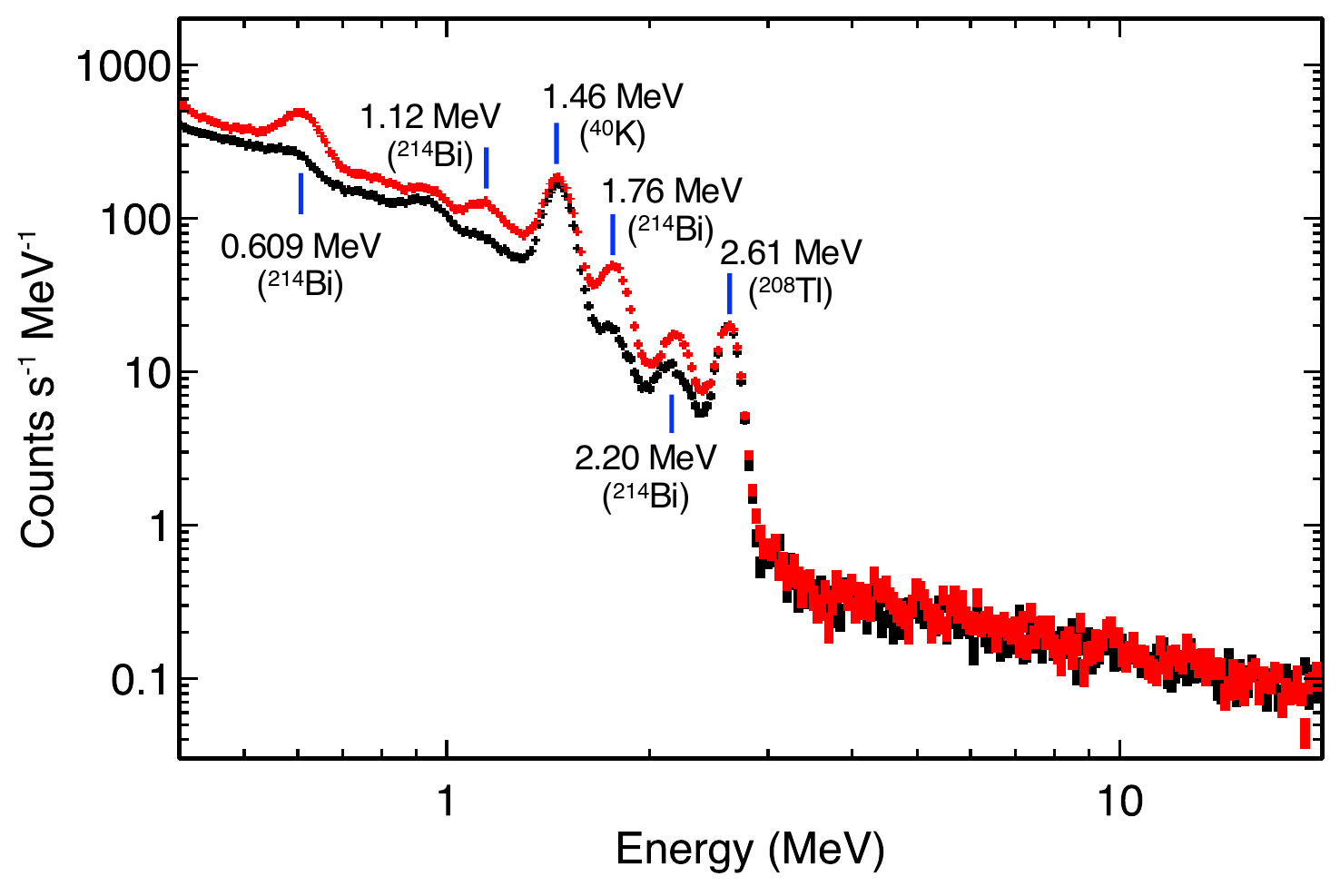}
	\caption{
	Figure 5: Averaged background spectra in 5th March 2018 (black; sunny) and 6th March 2018 (red; rainy) obtained at Kanazawa University.
	The spectra are accumulated for 30 minutes.}
	\end{center}
\end{figure}

\section{Results}
\subsection{Number of detected events}
The number of detected events since 2006 is presented in Figure~6.
	During 2006-2014, we operated only the Kashiwazaki site with 2 detectors.
	In average, 0.8 short bursts and 1.6 long bursts were observed for one winter season.
	Since 2015, we started the mapping observation campaigns.
	After the launch, the detection rate becomes 9.0 per year for long bursts, and 2.3 per year for short bursts.
	Among the 41 long bursts, we observe 5 events which abruptly terminated with lightning discharges.

\begin{figure}[ht]
	\begin{center}
	\includegraphics[width=\hsize]{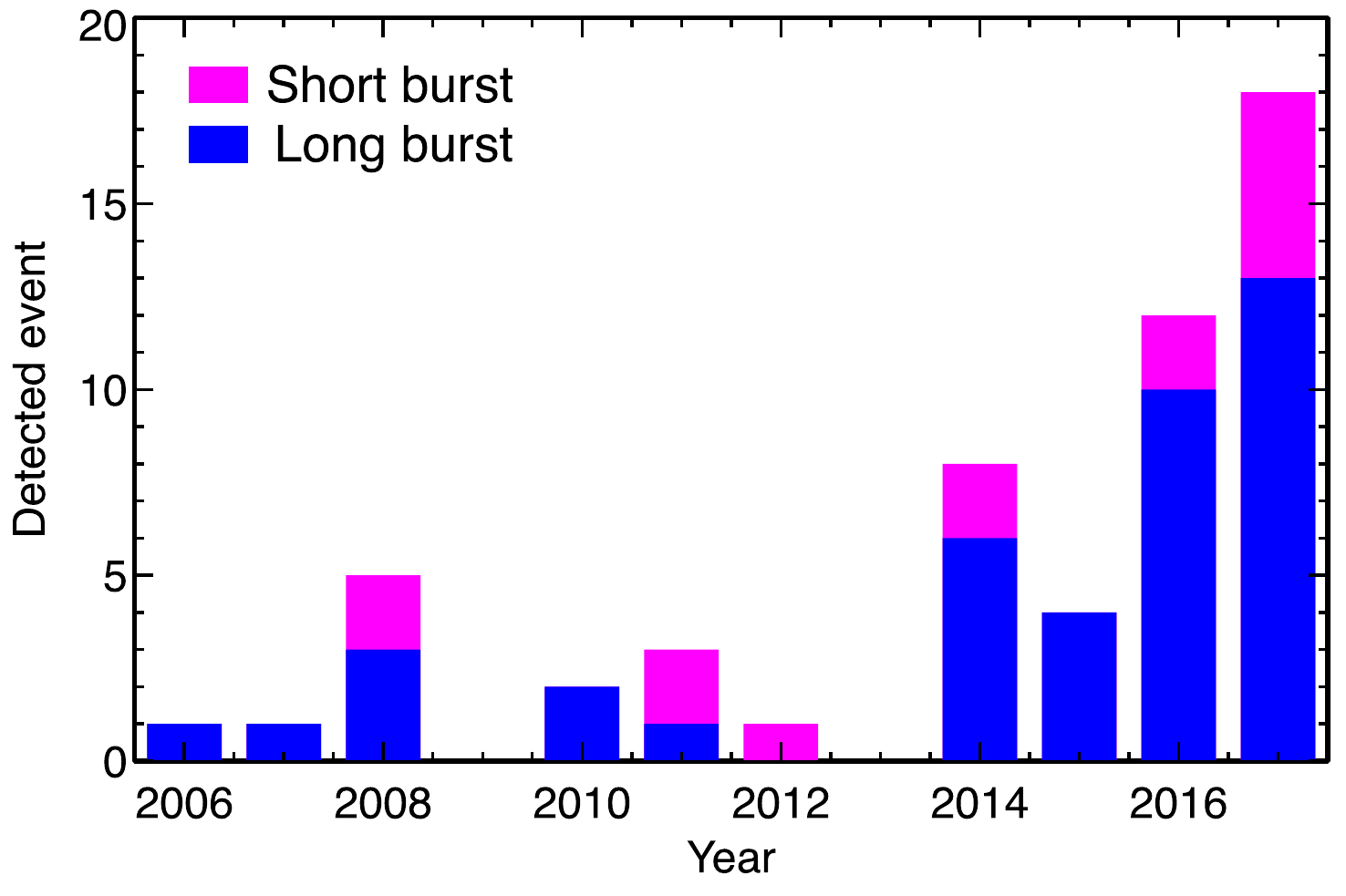}
	\caption{Figure 6: High-energy event detections from 2006-2007 to 2017-2018 winter seasons. 
	Blue and magenta bars show long bursts and short bursts, respectively.}
	\end{center}
\end{figure}

\begin{figure}[ht]
	\begin{center}
	\includegraphics[width=\hsize]{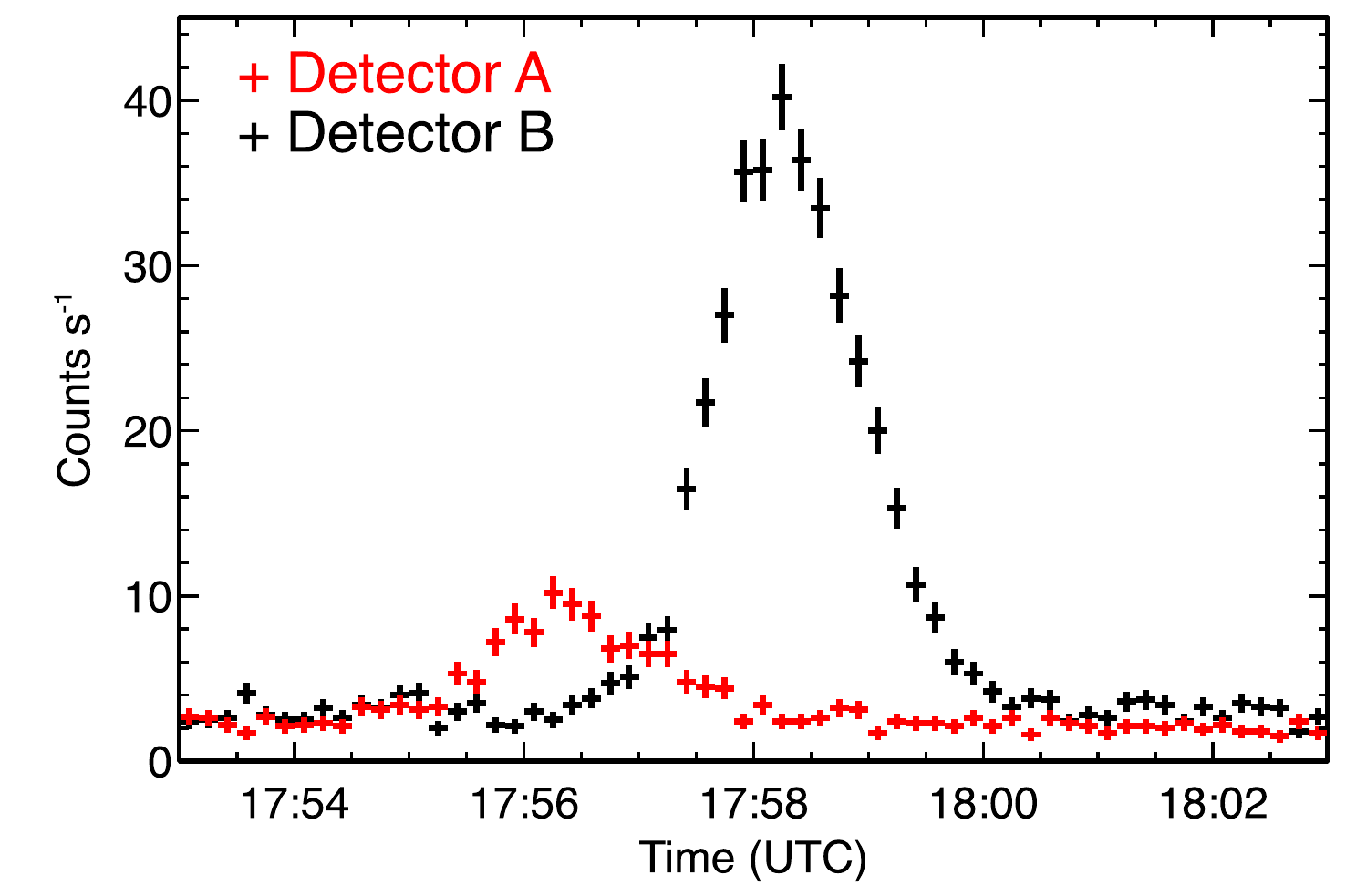}
	\caption{
	Figure 7: Count-rate histories of a long burst observed in the Komatsu sites with 10-sec bins.
	The energy range is 3.0--15.0~MeV. Red and black data points present count rates in detectors~A and B, respectively.
	}
	\end{center}
\end{figure}

\subsection{Tracking of long bursts}
One of the main purposes of the mapping observation project is to reveal life cycles of long bursts.
	We introduce a successful example to track an identical long burst by two radiation monitors.
	In the Komatsu site, two radiation monitors were deployed 
	at the roof of a high school and a science museum with 1.3~km separation.
	The monitors detected a long burst during heavy thunderstorms on 8th December 2016.
	Count-rate histories of the long burst are shown in Figure 7.
	We fitted the count-rate histories with a Gaussian function, and obtained count-rate peak time 
	at detectors A and B as 17:56:25.6~$\pm$~2.5 and 17:58:19.5~$\pm$~0.8 (UTC), respectively.
	Temporal separation of the peak time at two detectors is $114.0~\pm~2.6~{\rm sec}$. 

\begin{figure*}[hbt]
	\begin{center}
	\includegraphics[width=0.8\hsize]{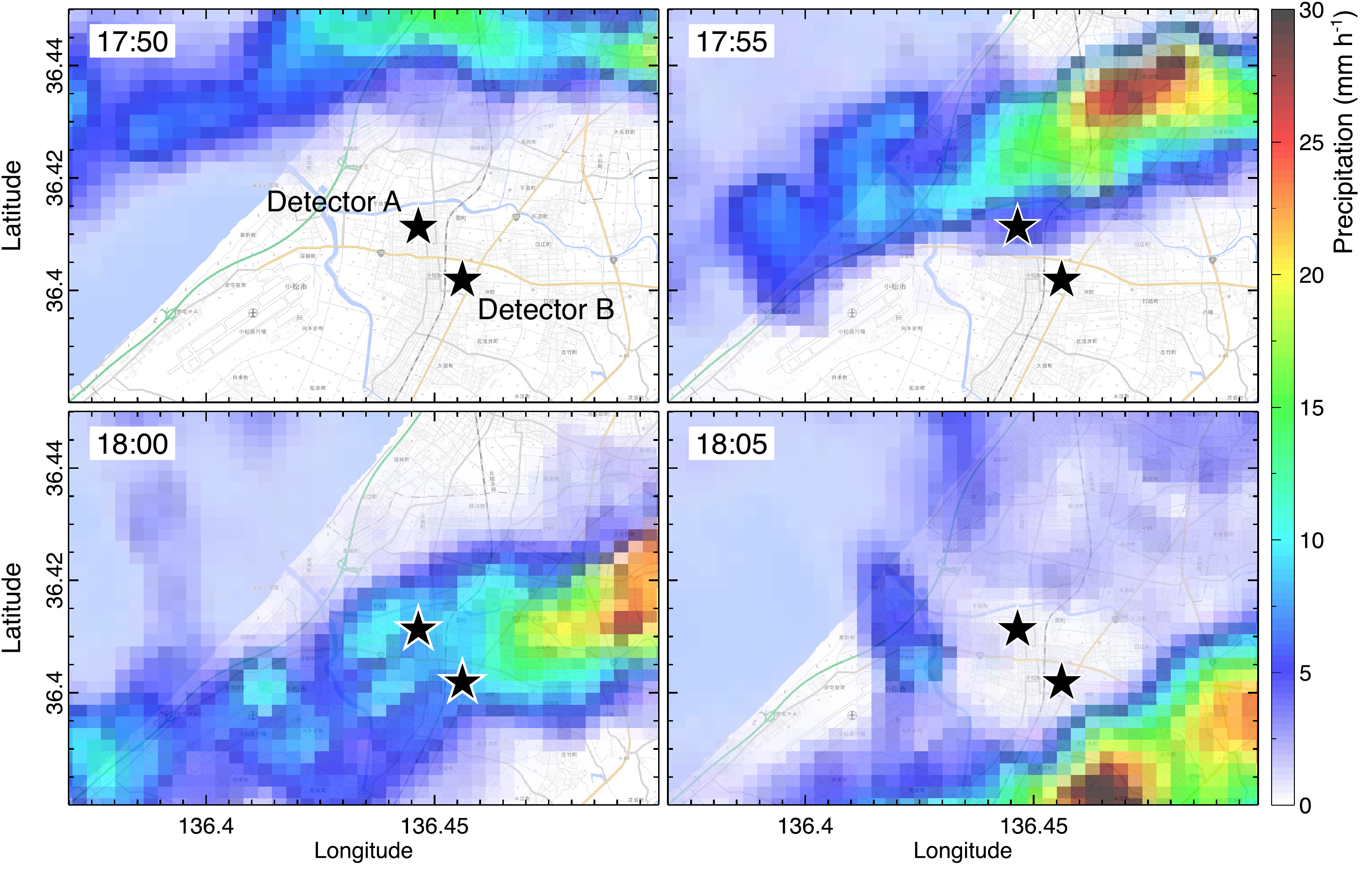}
	\caption{
	Figure 8: Five-minute interval precipitation maps in Komatsu obtained by XRAIN. 
	Black star markers present the observation sites.
	}
	\end{center}
\end{figure*}

To investigate the wind flow at that moment, we utilized data of XRAIN (eXtended RAdar Information Network).
	XRAIN is an X/C-band radar network operated by Japanese Ministry of Infrastructures, Land and Transportations.
	It can observe radar-echo and precipitation maps with a 1-minute interval.
	Precipitation maps with a 5-minute interval are shown in Figure 8.
	A high-precipitation area, namely thunderclouds, passed above the detectors from north-west to south-east
	during 17:55--18:00, consistent with the detection time of the long burst.
	In addition, it is indicated that detector~A located upwind was required to detect the long burst prior to detector B.
	This is also consistent with the detection order of the long burst.
	By comparing pairs of precipitation maps with the 5-minute interval, 
	we obtained the wind flowing at a speed of $10.9 \pm 1.2~{\rm m}~{\rm s}^{-1}$ with a direction of $296^{\circ}$. 

The expanded map in Komatsu is shown in Figure~9. The two observation sites has a 1.36~km separation.
	We assumed that the long burst moved with the wind flow.
	With the wind direction of $296^{\circ}$, the distance between the closest points 
	from the detectors to the center of the long burst is 1.20~km.
	The wind needed $110^{+14}_{-11}~{\rm sec}$ to pass the distance of 1.20~km.
	This is consistent with the temporal separation of the peak time at the detectors, $114.0~\pm~2.6~{\rm sec}$.
	Therefore, it is clear that an identical long burst moved with the ambient wind flow, then irradiated the two detectors with a time lag.

\begin{figure}[ht]
	\begin{center}
	\includegraphics[width=\hsize]{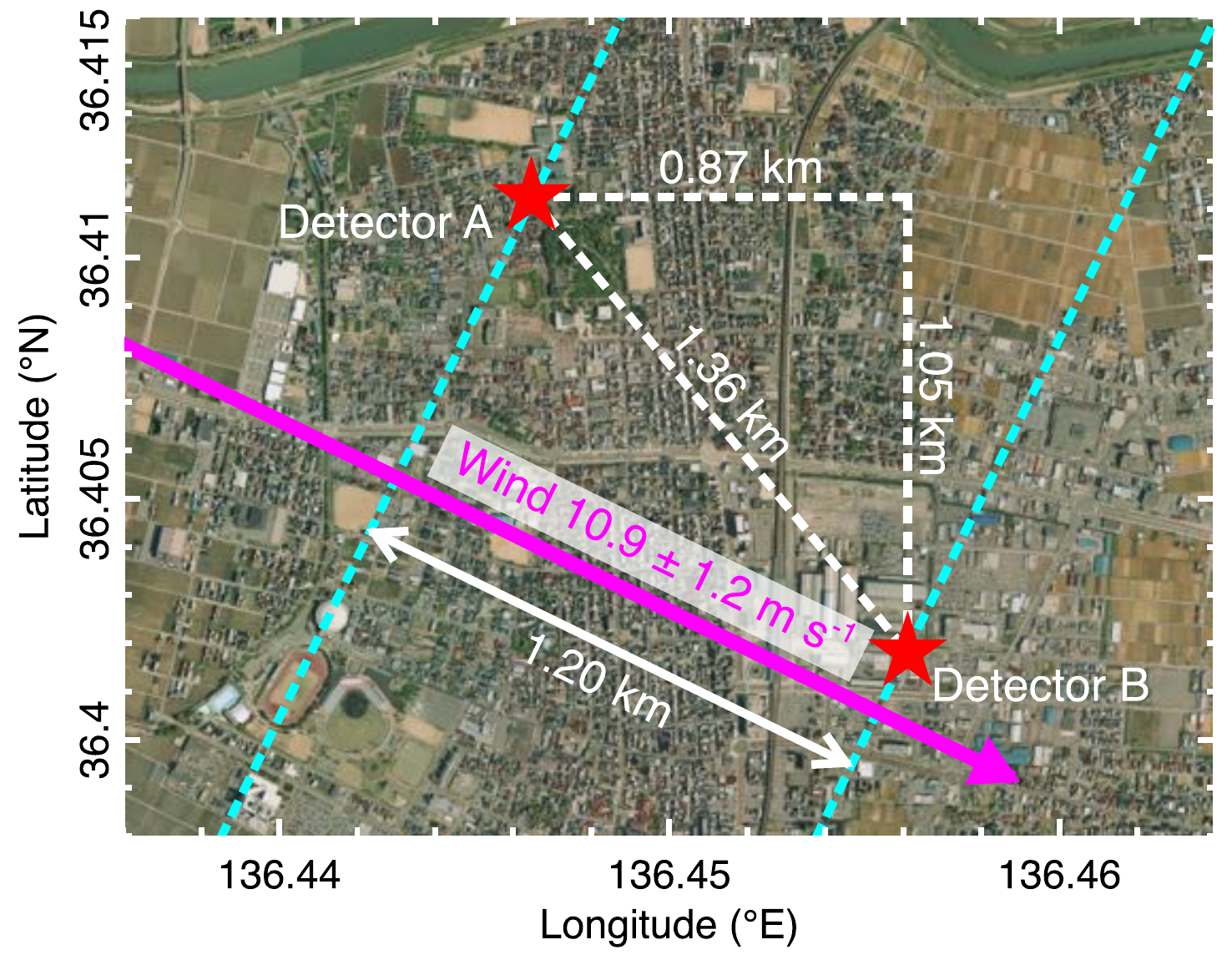}
	\caption{
	Figure 9: An aerial photograph in Komatsu. Red star markers and a magenta arrow present
	observation sites and wind direction, respectively.
	The cyan-dashed lines present possible areas of the long-burst center when the long burst reaches the closest position to each detector.
	}
	\end{center}
\end{figure}

In this case, the center of the long burst passed by detector~B closer than detector~A 
	because the peak count rate of detector B was higher than that of detector~B.
	On the other hand, the observation with two detectors cannot determine the exact position of the long burst center.
	Observations with 3 or more detectors will give us not only the burst center, but also structures of irradiated areas, total gamma-ray fluxes, and so on.
	In addition, time variations of long burst will be also revealed with the mapping observation project by tracking an identical long burst.
	In conclusion, our pilot observation in the 2016-2017 winter season demonstrated 
	a successful tracking of a long burst with multiple detectors, and suggested scientific importance of the mapping observation.

\subsection{Interpretation of short bursts}
In this section, we briefly summarize the interpretation of short bursts published as Enoto et al.\cite{Enoto_2017}
	The short burst event detected in the Kashiwazaki site enabled us 
	to demonstrate atmospheric photonuclear reactions triggered by a lightning discharge.
	On 6th February 2017, four detecters deployed in the Kashiwazaki site simultaneously detected a short burst
	coinciding with a lightning discharge reported by Japanese Lightning Detection Network.
	It lasted for $\sim$200~ms and decayed with a time constant of 50--60~ms.
	The spectra of the short burst present a continuum and sharp cutoff at 10~MeV, 
	which is different from bremsstrahlung (see Figure 1).
	At the beginning of the short burst, fast (less than  a few milliseconds) and large energy deposit 
	into scintillation crystals (more than hundreds of MeVs) were suggested by paralyzed output signals.
	After the short burst, two of the four detectors recorded an afterglow in the 0.4-0.6~MeV range lasting for $\sim$~1 minute. 
	The spectra apparently present the 0.511~MeV line of electron-positron annihilations.
	Since the annihilation emission was not accompanied by significant numbers of photons with energies more than 1 MeV, 
	this is not of the pair-production origin.
	
These are interpreted as photonuclear reactions triggered by a lightning discharge.
	First of all, the lightning discharge provoked the fast and large energy deposit at the beginning of the short burst, namely a downward TGF.
	Due to high-energy photons of more than 10~MeV, the downward TGF triggered atmospheric photonuclear reactions such as
	$^{14}{\rm N} + \gamma$ $\to$ $^{13}{\rm N} + n$ and $^{16}{\rm O} + \gamma$ $\to$ $^{15}{\rm O} + n$, producing fast neutrons. 
	These neutrons gradually lost their kinetic energy via elastic scattering in the atmosphere with time scale of $\sim$50~ms, and finally reacted with ambient $^{14}$N.
	Most of the neutrons were thought to exhibit charged-particle reaction $^{14}{\rm N} + n$ $\to$ $^{14}{\rm C} + p$, producing quasi-stable carbon isotope $^{14}$C.
	The rest reacted with $^{14}$N via neutron captures $^{14}{\rm N} + n$ $\to$ $^{15}{\rm N} + \gamma$.
	After the neutron captures, $^{15}$N immediately emitted de-excitation gamma rays consisting of multiple line emission.
	The short burst was caused by the de-excitation gamma rays. 
	The energy spectra are explained as superposition of such de-excitation gamma rays with moderate energy resolution of BGO scintillation crystals.
	The time scale of the short burst is consistent with that of neutron thermalization in the atmosphere.
	Such neutrons produced by a lightning discharge during winter thunderstorms have been also reported by Bowers et al.\cite{Bowers_2017}
	
The annihilation afterglow originate from the bi-products of photonuclear reactions $^{13}$N and $^{15}$O.
 	These isotopes emit positrons via beta-plus decay with decay constants of 10 and 2 minutes, respectively.
	The region where the photonuclear reactions were provoked is considered to be filled with $^{13}$N and $^{15}$O.
	Since the region can flow with the ambient wind, the detector recorded the annihilation emission only when the region was above it.
	Therefore, the duration of the annihilation emission is shorter than decay constant of the isotopes.

By photonuclear reactions, various isotopes such as $^{13}$N, $^{15}$O, $^{13}$C, $^{15}$O, and $^{14}$C are produced.
	Our result thus demonstrates a new channel for isotope production inside the Earth's atmosphere. 
	Especially, $^{14}$C is the important isotope for dating method of archeology.
	Therefore, how many the $^{14}$C isotopes are produced by lightning is of great importance.
	In addition, what type of lightning can trigger TGFs and photonuclear reactions still remains as an open question.
	Our further observation will answer the questions, as well as give indications on questions for ordinal TGFs observed from space.

\subsection{Termination of long bursts}
In this section, the observation reported in Wada et al.\cite{Wada_2018} is presented.
	We performed an observation campaign at the Suzu site with high-energy radiation and atmospheric electric field (AEF) monitors.
	This observation site was also monitored by a low-frequency lightning mapping network (LF network).
	The LF network consists of 5 stations installed along Toyama Bay, which has a flat plate antenna sensitive to the 0.8--500~kHz radio frequency band.

On 11th February 2017, the gamma-ray monitors recorded a long burst lasting for $\sim$1~minute
	as the AEF monitor detected a negatively-charged thundercloud approaching.
	The energy spectrum of the long burst extends up to 20~MeV, 
	and is well reproduced by a power-law function with an exponential cutoff.
	When it was reaching its maximum flux, the long burst was abruptly terminated.
	At that moment, the AEF monitor detected a pulse indicating a lightning discharge.

The lightning discharge was also detected by the LF network.
	The LF network recorded continuous waveform lasting for 300~ms.
	Most pulses of the waveform are small-amplitude emissions such as stepped leaders, thus originate from a leader development.
	The lightning discharge was initiated 15~km west from the gamma-ray detectors, then developed for 300~ms with 70~km expansion.
	Several ones of the lightning pulses were located within 1~km from the radiation detectors.
	Since timing of the pulses close to the observation site is consistent to the moment when the long burst was terminated, 
	we conclude that the long burst was terminated by the leader development.
	
This is the first simultaneous detection of the long burst termination with gamma-ray, AEF, and LF mapping observation.
	It proves that the combination of these methods gives us the clues to understand the mechanism of long bursts.
	As the collaboration continues for observation in Japanese winter thunderstorms, 
	it will provide us new sites into the phenomena.

\section{Conclusion}
We launched the mapping observation project for high-energy phenomena in Japanese winter thunderstorms in 2015.
	The portable gamma-ray detectors with the new DAQ system dedicated to the project were developed, 
	and up to 16 detectors were deployed and operated in coastal areas of Japan Sea during 2015-2018 winter seasons.
	During the three-year observations, we detected 27 long bursts and 8 short bursts.
	The number of detected events increases as more detectors are deployed.
	
Owing to the mapping observation, we succeeded in observing the identical long burst 
	moving with ambient wind flow by using 2 gamma-ray detectors.
	Also, the observation in the Kashiwazaki site with 4 detectors enabled us to interpret the short burst 
	as atmospheric photonuclear reactions triggered by the TGF at ground level.
	In addition, the collaborative campaign in Suzu with the AEF and LF measurements
	allowed us to investigate the relation between lightning and long bursts, 
	and the charge structure responsible for electron acceleration.
	We demonstrated that the mapping observation project continues to give us fruitful scientific results.
	Further observations not only with gamma-ray but also radio-band and electric-field measurements
	will enable us to resolve the questions in high-energy atmospheric physics.



%


\section{Acknowledgement}
We deeply thank M. Kamogawa, G. S. Bowers and D. M. Smith for the collaborative observation in Suzu, 
	T. Morimoto and Y. Nakamura for LF observation, M. Sato and Y. Sato for interpretation of the short burst event, 
	S. Otsuka, H. Kato, and T. Takagaki for detector development.
	Detector deployment was supported by D. Yonetoku, T. Sawano, K. Watarai, K. Yoneguchi, K. Kimura, K. Kitano, 
	K. Kono, K. Aoki, staffs of Kanazawa University Noto School, 
	and the radiation safety group of Kashiwazaki Kariwa Nuclear Power Plant, Tokyo Electric Power Company Holdings.
	This work is supported by JSPS/MEXT KAKENHI grants 15K05115,15H03653, 16H06006, 18J13355,
	by Hakubi project and SPIRITS 2017 of Kyoto University, 
	and by the joint research program of the Institute for Cosmic Ray Research (ICRR), the University of Tokyo.
	The background images in Figure 4, 8 and 9 were provided by the Geospatial Information Authority of Japan.
	The XRAIN data obtained by Japan Ministry of Land, Infrastructure, Transport and Tourism 
	was retrieved from Data Integration and Analysis System (DIAS) operated by the University of Tokyo.



\end{document}